\documentclass[letterpaper]{JHEP3}

\usepackage{amssymb}
\usepackage{amsmath}
\usepackage{bbold}
\usepackage{graphicx}
\usepackage{eufrak}

\usepackage[hang,flushmargin]{footmisc} 

\allowdisplaybreaks[4]

\def\bea{\begin{eqnarray}}
\def\eea{\end{eqnarray}}
\def\beq{\begin{equation}}
\def\eeq{\end{equation}}
\def\nn{\nonumber}
\newcommand{\ignore}[1]{}

\def\<{\langle}
\def\>{\rangle}

\def\ot{\overline{\tau}}

\def\Nup{N_{\rm up}}

\title{Multiverse rate equation including bubble collisions}

\author{Michael P.~Salem\\
Stanford Institute for Theoretical Physics and Department of Physics, 
Stanford University,\\ Stanford, CA 94305, USA}

\abstract{The volume fractions of vacua in an eternally inflating multiverse are described by a coarse-grain rate equation, which accounts for volume expansion and vacuum transitions via bubble formation.  We generalize the rate equation to account for bubble collisions, including the possibility of classical transitions.  Classical transitions can modify the details of the hierarchical structure among the volume fractions, with potential implications for the staggering and Boltzmann-brain issues.  Whether or not our vacuum is likely to have been established by a classical transition depends on the detailed relationships among transition rates in the landscape.}

\preprint{SU-ITP-12/35}

\begin{document}

\section{Introduction}
\label{sec:introduction}

We might live in an eternally inflating multiverse.  Eternal inflation occurs whenever a sufficiently large volume is in a state sufficiently close to a vacuum in which the energy density is positive and the decay rate is smaller than the Hubble rate \cite{Guth:1982pn}, and/or whenever a sufficiently homogeneous field configuration evolves in a sufficiently flat, positive interaction potential \cite{Vilenkin:1983xq,Linde:1986fd}.  These statements contain a number of qualifications, so it is worth noting that the first set of conditions is satisfied by the observed state of the local universe, if the so-called dark energy is due to vacuum energy, while the second set is similar to the initial conditions implicit in the simplest models of slow-roll inflation, though with the inflaton further up its potential \cite{Guth:2000ka}.  When it occurs, eternal inflation generates an endless spacetime in which every phase of vacuum takes place in a fractal mosaic of widely separated domains \cite{Aryal:1987vn,Garriga:1997ef}, the various vacua being attained either by bubble formation \cite{Coleman:1980aw,Brown:2007sd}, by stochastic diffusion \cite{Goncharov:1987ir}, and/or by other processes \cite{Hawking:1981fz,BlancoPillado:2009di,Easther:2009ft,Balasubramanian:2010kg,Brown:2011ry,Kleban:2011cs}.  

In an eternally inflating multiverse, the fraction of the global spacetime volume occupied by any of the various vacua is not directly observable.  Nevertheless, these volume fractions might be relevant to understanding the local conditions in our universe.  For instance, the proper theory of initial conditions might be a theory of the multiverse as a whole, with our local ``initial'' conditions---i.e.~the conditions describing the onset of slow-roll inflation in the particular phase of vacuum that gives rise to our universe---being selected according to their prevalence in the global spacetime.  It is also possible that these volume fractions express a holographic dual to bulk spacetime physics \cite{Garriga:2008ks,Garriga:2009hy,Bousso:2009dm,Bousso:2010id,Vilenkin:2011yx,Harlow:2011az}.  

In a multiverse where vacuum transitions occur predominantly via bubble formation, the volume fractions occupied by the various vacua are described by a rate equation \cite{Garriga:1997ef,Garriga:2005av}.  An important caveat is that the predictions of the rate equation depend on the choice of global time foliation.  This is because different global time foliations explore different regions in the diverging spacetime at different rates, while the exponential expansion of eternal inflation ensures that most of the total volume is near the boundary at any finite time cutoff.  This expresses the measure problem of eternal inflation (for some recent reviews, see for example \cite{Vilenkin:2006xv,Guth:2007ng,Freivogel:2011eg,Salem:2011qz})  Resolving the measure problem is of fundamental importance to (eternal) inflationary cosmology, but it is tangential to the thrust of this paper. 

This paper concerns another shortcoming of the rate equation, which is that it ignores bubble collisions.  Semiclassical vacuum transition rates are exponentially suppressed, and a contribution to the rate equation from a bubble collision should involve a product of two such rates (one for each bubble in the collision), and so one might argue that bubble collisions can be ignored by expanding in powers of transition rates.  However, these rates are exponentially staggered, meaning the product of two transition rates could be much larger than another transition rate.  Moreover, over the course of its evolution each bubble collides with a diverging number of other bubbles.  This divergence is regulated by the aforementioned measure, but it is not a priori clear how this resolution will play out.  

Although the detailed phenomenology of the rate equation is rather technical, for those very familiar with the literature (including the standard notation and assumptions) our conclusions are simple to state.  (Those less familiar with the literature will find the conclusions of this paragraph explained more thoroughly in the main text.)  After including the effects of bubble collisions, the rate-equation transition matrix becomes
\beq
M_{ij} = \kappa_{ij} - \delta_{ij}\kappa_i+\sum_{k,\ell}\gamma_{ik\ell j}\kappa_{\ell j}\kappa_{kj}\,,
\label{Mabs}
\eeq 
where $\gamma_{ik\ell j}$ is related to the average volume fraction in vacuum $i$ in the causal futures of collisions between bubbles of vacua $k$ and $\ell$, when these bubbles nucleate in vacuum $j$.  The effects of bubble collisions are most significant when the first two terms in (\ref{Mabs}) are zero but (because of classical transitions) the third term is not.  Meanwhile, the components of the dominant eigenvector $s_i$ of $M_{ij}$ are, to leading order,   
\beq
s_i = \sum_{\{p_a\}}\,\frac{\kappa_{ip_1}\!+\sum_{j,k}\gamma_{ijkp_1}\kappa_{jp_1}\kappa_{kp_1}}{\kappa_i-q} 
\times\ldots\times\frac{\kappa_{p_n1}+\sum_{j,k}\gamma_{p_njk1}\kappa_{j1}\kappa_{k1}}{\kappa_{p_n}\!-q} \,,
\label{intro}
\eeq
where the sum covers the sequences of transitions that connect the dominant vacuum ``1'' to the vacuum $i$ using the fewest number of upward transitions (and ``leading order'' refers to an expansion in these upward transition rates).  Again, the effects of classical transitions can be significant because the first term in each factor can be zero when the second term is not.  Note that the dominant vacuum---defined as the positive-energy vacuum with the smallest decay rate---still dominates the volume fraction.  In particular, even if the dominant vacuum can set up classical transitions to another positive-energy vacuum, the volume fraction of the latter is still much less than unity.  On the other hand, the detailed hierarchical structure among the components of $s_i$ can be modified by the existence of classical transitions.   

The detailed hierarchical structure among the volume fractions of the various vacua in the landscape are relevant to the so-called staggering issue, which concerns the competition between anthropic selection for small (magnitude) vacuum energies and cosmological selection for large volume fractions when attempting to explain the observed size of the cosmological constant; see for example \cite{SchwartzPerlov:2006hi,SchwartzPerlov:2006hz,Olum:2007yk,SchwartzPerlov:2008he}.  It is also relevant to the so-called Boltzmann-brain issue, which concerns the likelihood for observers to arise in an extremely low-entropy Hubble volume such as we observe, as opposed to in a relatively high-entropy Hubble volume such as describes the distant future \cite{Dyson:2002pf,Albrecht:2004ke,Page:2006dt,Bousso:2006xc,Bousso:2011aa}.  One might also take interest in the likelihood that our vacuum was created by a classical transition, as opposed to by semiclassical bubble formation.  Although we discuss these issues, a conclusive investigation requires a detailed understanding of the landscape, and is beyond the scope of this paper.       

The remainder of this paper is organized as follows.  In Section \ref{sec:rate1} we review the construction of the rate equation, ignoring bubble collisions, while in Section \ref{sec:coll} we include bubble collisions.  We predominantly work in terms of a scale-factor-time foliation, though we briefly explain how to translate the results into those of a lightcone-time foliation.  The phe\-nom\-e\-nol\-ogy of the rate equation is studied in Section \ref{sec:phenomenology}.  We begin with a review of a simple toy landscape, ignoring bubble collisions, and then we explore the toy landscape while including the effects of some representative classical transitions.  We extend our results to a more general landscape in Section \ref{ssec:generalize}, where we briefly discuss the staggering issue.  In Section \ref{ssec:BBs} we discuss the Boltzmann-brain issue and in Section \ref{ssec:past} we discuss the likelihood of a classical transition in our past.  Finally, we draw our conclusions in Section \ref{sec:conclusions}.

\section{Rate equation without bubble collisions}
\label{sec:rate1}

To begin, we reconstruct the standard rate equation, adopting the usual assumptions \cite{Garriga:1997ef,Garriga:2005av}.  In particular, we take the spacetime to be everywhere (3+1)-dimensional (the rate equation in a transdimensional multiverse is studied in \cite{SchwartzPerlov:2010ne}), we assume that all vacuum transitions occur via semiclassical bubble formation, we coarse-grain over the time scales of any transient cosmological evolution between epochs of vacuum-energy domination, we assume there are no vacua with precisely zero vacuum energy,\footnote{The scale-factor and lightcone-time cutoff measures both fail to regulate the volume in bubbles with precisely zero vacuum energy.  (The author thanks Adam Brown for explaining the problem with the scale-factor cutoff measure.)  Note that four-dimensional vacua with precisely zero vacuum energy might not exist.} and we ignore bubble collisions.  In the next section we include bubble collisions, but the other assumptions are held throughout the analysis.  

The rate equation describes the volume fractions of vacua in terms of some global time variable $t$.  To establish the global time foliation, we start with a large, spacelike hypersurface $\Sigma_0$ on which we set $t=0$.  Note that it is not necessary for $\Sigma_0$ to be a Cauchy surface: if $\Sigma_0$ intersects an eternal worldline (that is, a worldline that never sees a vacuum energy density less than zero), then the rate equation describing the future evolution of $\Sigma_0$ will possess an attractor solution that is independent of the detailed orientation and distribution of vacua on $\Sigma_0$.  This implies that the volume fractions are independent of the choice of $\Sigma_0$.

Suppose the total physical volume of $\Sigma_0$ in vacuum $i$ is $V_i$.  To construct the rate equation, we first compute the change in physical volume in vacuum $i$ over a time interval $\Delta t$, i.e.  
\beq
\Delta V_i = V_i(\Sigma_{\Delta t})-V_i(\Sigma_0) \,,
\eeq
where $\Sigma_{\Delta t}$ denotes the hypersurface of constant $t=\Delta t$.  The coarse-grain approach explores the limit of large $\Delta t$ to compute the various contributions to $\Delta V_i$, but then expands in $\Delta t\ll 1$ to construct a differential equation.  In the context of bubbles with positive vacuum energy densities---henceforth referred to as dS bubbles---this means that the rate equation ignores the transient cosmological evolution between epochs of vacuum energy domination.  In the context of bubbles with negative vacuum energy densities---henceforth referred to as AdS bubbles---it means the rate equation ignores the cosmological evolution altogether.\footnote{\label{foot1}This does not imply that the rate equation cannot be used to study cosmological evolution on time scales that are smaller than the coarse-graining, in dS or AdS vacua.  The rate equation gives the bubble nucleation rate for all vacua that are reached primarily via semiclassical bubble nucleation starting from a dS vacuum.  The bubble nucleation rate can then be combined with a more fine-grained analysis to study the dynamics in bubbles on smaller time scales.  Some examples are reviewed in the appendix of \cite{Salem:2011qz}.}  Although the coarse-grain rate equation does not by itself provide an accurate assessment of the physical volume fractions in AdS vacua, it is convenient to track the volume fractions in these vacua anyway.  We do this by simply conserving comoving volume during transitions to AdS vacua, and ignoring the subsequent evolution of the volume.

\subsection{Scale-factor time}

We first take $t$ to be the scale-factor time \cite{Linde:1993xx,DeSimone:2008bq,Bousso:2008hz,DeSimone:2008if,Bousso:2012tp},  
\beq
dt = H d\tau \,,
\eeq
where $\tau$ is the proper time evaluated along a geodesic congruence orthogonal to $\Sigma_0$, and $H$ is the local Hubble rate, in particular we can take $H\equiv (1/3)\,u^\mu_{\phantom{\mu};\mu}$ in terms of the four-velocity field $u^\mu$ along the congruence.  Although a precise definition of scale-factor time involves some subtleties in the treatment of locally contracting spacetime regions \cite{Bousso:2008hz,DeSimone:2008if,Bousso:2012tp}, these can be ignored in the coarse-grain analysis, which smears over such regions.  
  
\begin{figure}[t!]
\begin{center}
\includegraphics[width=0.4\textwidth]{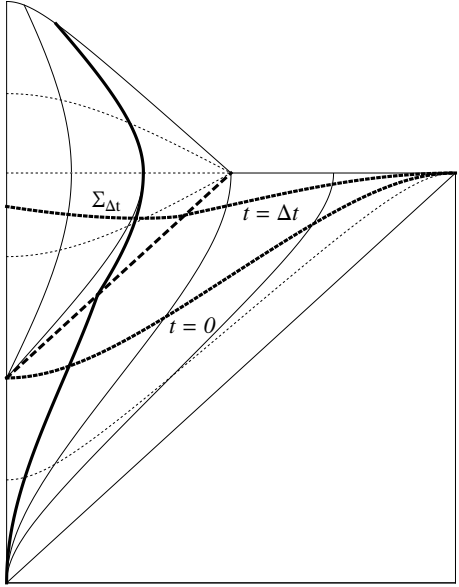}
\caption{\label{fig:matching}Toy conformal diagram of (dS) bubble formation.  The bubble wall asymptotes to the thick dashed line.  Outside of the bubble, curves indicate surfaces of constant $r$ (solid) and $\tau$ (dotted) in the spatially flat chart.  Surfaces of constant $\tau$ are also surfaces of constant scale-factor time $t$; the surfaces  $t=0$ and $t=\Delta t$ are illustrated (see main text).  Inside the bubble, curves indicate surfaces of constant $\xi$ (solid) and $\ot$ (dotted) in the open FRW chart.  The thick solid curve indicates a geodesic that is initially comoving with respect to the spatially flat chart which enters the bubble and soon becomes comoving with respect to the open FRW chart.  The thick dotted line marked $\Sigma_{\Delta t}$ continues the surface of constant scale-factor time $t=\Delta t$ into the bubble.}
\end{center}
\end{figure}

Before proceeding, it is helpful to collect some facts about bubble formation.  Consider a region in some dS vacuum $i$ with cosmological constant $\Lambda_i$.  On scales that are small compared to the curvature radius, the line element can be written
\beq
ds^2 = -d\tau^2 + \frac{1}{4H_i^2}\,e^{2H_i\tau}\left( dr^2+r^2\,d\Omega^2 \right) ,
\eeq  
where
\beq
H_i \equiv \sqrt{|\Lambda_i|/3} \,,
\eeq
$d\Omega^2$ is the line element on the unit 2-sphere, and the absolute value is inserted to establish a general definition for when we consider negative values of $\Lambda_i$.  Now suppose an initially point-like bubble nucleates at time $\tau=\tau_{\rm nuc}$.  To facilitate future reference we take the hypersurface $\tau=\tau_{\rm nuc}$ to coincide with the aforementioned $\Sigma_0$, at least in the vicinity of the bubble.  (A diagram is provided in Figure \ref{fig:matching}.)  The bubble wall of a point-like bubble expands at the speed of light.  Therefore, the comoving radius of the bubble at times $\tau\geq\tau_{\rm nuc}$ is
\beq
r_{\rm w}(\tau) = 2\,e^{-H_i\tau_{\rm nuc}}-2\, e^{-H_i\tau} \,.
\eeq
The bubble expands so as to subtend a finite comoving volume $(32\pi/3)\,e^{-3H_i\tau_{\rm nuc}}$ in the limit $\tau\to\infty$.  Note that this comoving volume corresponds to a physical volume $(4\pi/3)\,H_i^{-3}$ on the hypersurface $\Sigma_0$.  Therefore, the loss of physical volume in vacuum $i$ in the future evolution of $\Sigma_0$ due to the nucleation of this bubble is equivalent (in the limit $\tau\to\infty$) to the loss of physical volume that results from simply ignoring the would-be future evolution of a physical volume $(4\pi/3)\,H_i^{-3}$ on the hypersurface $\Sigma_0$.  

Now suppose that this bubble is a dS bubble.  The coarse-grain analysis ignores dynamics on time scales smaller than the time scale of vacuum domination; therefore the line element in the bubble can generically be written
\beq
ds^2 = -d\ot^2 + \frac{1}{4H_j^2}\,e^{2H_j\ot}
\left[d\xi^2 + \sinh^2(\xi)\,d\Omega^2\right]\,,
\label{bubblemetric}
\eeq
where $j$ labels the vacuum in the bubble, and it is implicit that we focus on bubble FRW times $\ot\gg H_j^{-1}$.  Importantly, surfaces of constant FRW time $\ot$ in the bubble are not surfaces of constant scale-factor time $t$ in the global foliation.  In particular, it can be shown that the change in scale-factor time between the hypersurface $\Sigma_0$ (corresponding to $\tau=\tau_{\rm nuc}$) in vacuum $i$ and a hypersurface of constant $\ot$ in the bubble is \cite{Bousso:2008hz}
\beq
\Delta t = H_j\ot + \ln(H_i/H_j) -\ln(2) + \xi
+\frac{2}{3}\ln\!\left(\frac{1}{2}+\frac{1}{2}e^{-\xi}\right) \,.
\label{Dt}
\eeq 
Therefore, the induced metric on $\Sigma_{\Delta t}$ when it overlaps with the bubble is
\bea
ds^2 &=& \left\{ \frac{1}{H_i^2}\,e^{2\Delta t}
\left(\frac{1}{2}+\frac{1}{2}e^{-\xi}\right)^{\!\!-4/3}\!e^{-2\xi}
-\frac{1}{H_j^2}\left[1-\frac{2}{3}
\left(1+e^\xi\right)^{-1}\right]^2\right\} d\xi^2 \nn\\
& & +\, \frac{1}{H_i^2}\,e^{2\Delta t}
\left(\frac{1}{2}+\frac{1}{2}e^{-\xi}\right)^{\!\!-4/3}\! e^{-2\xi}\, 
\sinh^2(\xi)\,d\Omega^2 \,. 
\label{induced}
\eea
The coarse-grain approximation explores the limit where the first term in brackets dominates over the second (because the coarse-grain approximation implies for example $\Delta t > H_i/H_j$).  Accordingly, the physical three-volume of the intersection of $\Sigma_{\Delta t}$ and the bubble is
\beq
\int_0^\infty d\xi\, 4\pi\sinh^2(\xi)\,\frac{1}{H_i^3}\, e^{3\Delta t}
\left(\frac{1}{2}+\frac{1}{2}e^{-\xi}\right)^{\!\!-2}e^{-3\xi}
= \frac{4\pi}{3H_i^3}\,e^{3\Delta t} \,.
\label{vol1}
\eeq
Technically, the integrand in (\ref{vol1}) is invalid at large values of $\xi$, for which the hypersurface $\Sigma_{\Delta t}$ explores times $\ot\lesssim H_j^{-1}$ and the above results receive corrections.  Nevertheless, for sufficiently large values of $\Delta t$ the physical volume on $\Sigma_{\Delta t}$ is dominated by regions for which $\ot\gg H_j^{-1}$, and the above approximations are accurate.  

Recall that we modeled the initial bubble as point-like.  A realistic bubble has some nonzero initial radius, and the bubble wall has zero initial velocity.  However, the bubble wall accelerates, its velocity approaching the speed of light.  Therefore, in the limit of large $\Delta t$, our results for point-like initial bubbles coincide with the results for more realistic bubbles.  An important exception to this rule is an ``upward'' transition from lower to higher vacuum energy, for which the bubble wall fills the horizon \cite{Brown:2007sd}.  Lacking any clearer guidance, we simply use the above results for these vacua as well.

We now return to the rate equation.  Breaking the calculation of $\Delta V_i$ up into parts, we first compute the change in physical volume in comoving regions that begin and remain in vacuum $i$ over the interval $\Delta t$.  This is given by the definition of scale-factor time:
\beq
\Delta V_i = \left( e^{3\Delta t}-1\right) V_i \,\to\, 3V_i\,\Delta t \,,
\label{exp}
\eeq
for any dS vacuum $i$.  As mentioned above, we simply ignore the change in physical volume in comoving regions that begin and remain in the same AdS vacuum, as these vacua collapse into a big-crunch singularity on time scales that are small compared to the coarse-graining.  Hence, for these vacua we take $\Delta V_i=0$.  

Next, we compute the change in physical volume due to vacuum decay in comoving regions that begin in vacuum $i$.  Referring to the above analysis, we note that when $i$ is a dS vacuum then the effect of such decays is equivalent to removing physical volume $(4\pi/3)\,H_i^{-3}$ for each bubble nucleation, at the time of its nucleation.  The number of such nucleations in an interval $\Delta t$ is equal to the four-volume in vacuum $i$ in that interval times the decay rate,
\beq
\Delta N_{\rm decay} = \sum_j\Gamma_{ji}V_i\,\Delta\ot(\Delta t) = 
\sum_j\frac{\Gamma_{ji}}{H_i}V_i\, \Delta t \,, 
\label{dnumber}
\eeq
where $\Gamma_{ji}$ denotes the transition rate from vacuum $i$ to vacuum $j$, per unit physical three-volume per unit proper time, and we have used $\Delta t = H_i\,\Delta\ot$ in vacuum $i$.  The corresponding change in volume in vacuum $i$ is therefore
\beq
\Delta V_i = - \sum_j\frac{4\pi\Gamma_{ji}}{3H_i^4}V_i\,\Delta t \,.
\label{dV1}
\eeq
This result assumes that $i$ is a dS vacuum.  Although AdS vacua can also decay \cite{Harlow:2010az}, when they do so they transition exclusively to other AdS vacua, and therefore in the coarse-grain approach these transitions can be ignored (i.e.~we set $\Gamma_{ij}=0$ for AdS $j$).

Finally, we compute the change in physical volume due to transitions to vacuum $i$ from comoving regions that begin in some other vacuum.  The number of such transitions is computed in analogy to (\ref{dnumber}), but now with reference to the transition rate from some (dS) vacuum $j$ to vacuum $i$, summing over the relevant vacua $j$.  In the limit of large $\Delta t$, each such transition generates a physical volume (\ref{vol1}) in vacuum $i$, when $i$ is a dS vacuum.  This can be interpreted as the immediate creation of a physical volume $(4\pi/3)\,H_j^{-3}$, times a growth factor $e^{3\Delta t}$ which is already accounted for in (\ref{exp}).  Thus we write
\beq
\Delta V_i = \sum_j\frac{4\pi\Gamma_{ij}}{3H_j^4}\,V_j\,\Delta t \,,
\label{dV2}
\eeq
for a dS vacuum $i$.  For AdS vacua $i$, the analysis surrounding (\ref{vol1}) no longer applies.  Note however that the physical volume created in $i$ due to transitions from a given dS vacuum $j$ in (\ref{dV2}) is precisely the same as the physical volume lost in $j$ due to transitions to a given vacuum $i$ in (\ref{dV1}) (which requires exchanging the indices $i\leftrightarrow j$).  Since both of these expressions have the volume expansion factors stripped away, this equivalence expresses the conservation of comoving volume in vacuum transitions.  Extending this principle to AdS vacua indicates that we should use (\ref{dV2}) when describing the creation of volume in vacua $i$ for dS and AdS vacua $i$.  We make an error for AdS vacua since we ignore transitions from one AdS vacuum to another; however since these transition rates are exponentially suppressed next to the time scales of the AdS big crunches, this error is small.      

Combining results and taking the infinitesimal limit, we obtain
\beq
\frac{dV_i}{dt} = 3V_i + \sum_j\kappa_{ij}V_j - \sum_j\kappa_{ji}V_i \,,
\label{r10}
\eeq
where the first term is understood to apply only when $i$ is a dS vacuum, and we have defined the dimensionless decay rates 
\beq
\kappa_{ij}\equiv \frac{4\pi\Gamma_{ij}}{3H_j^4}.
\label{kdef}
\eeq
It is convenient to also define the volume fractions $f_i$, where
\beq
f_i \equiv \frac{V_i}{\sum_j V_j} \,.
\label{fidef}
\eeq
Since the dS vacua dominate the physical volume in the future evolution of $\Sigma_0$, we can write $\frac{d}{dt}\sum_jV_j=3\sum_jV_j$, which means the rate equation can be written
\beq
\frac{df_i}{dt} = M_{ij}f_j \,, 
\label{r10b}
\eeq
where the transition matrix is $M_{ij}\equiv \kappa_{ij}-\delta_{ij}\sum_k\kappa_{ki}$.  The solution to (\ref{r10b}) is \cite{Garriga:2005av}
\beq
\begin{array}{lll}
\phantom{hhhh} & \displaystyle f_i(t) \propto f^{(0)}_i+s_i\,e^{-qt} + \ldots 
& \quad\mbox{(anti--de Sitter)} \phantom{\Big[\Big]}\\
\phantom{hhhh} & \displaystyle f_i(t) \propto s_i\,e^{-qt} + \ldots 
& \quad\mbox{(de Sitter)} \,, \phantom{\Big[\Big]}
\end{array}
\label{ratesol}
\eeq
where the $f_i^{(0)}$ are constants reflecting the initial conditions, $q>0$ is (minus) the smallest-magnitude eigenvalue of $M_{ij}$, $s_i$ is the corresponding eigenvector---called the dominant eigen\-vector---and the ellipses denote terms that fall off faster than $e^{-qt}$.  Although we have presented the solutions for both dS and AdS vacua, as we have remarked the coarse-grain rate equation does not reliably assess the volume fractions in the AdS vacua.

\subsection{Lightcone time}
\label{ssec:lct}

The above analysis is sensitive to the choice of scale-factor time as the global time parameter $t$.  Another popular choice is lightcone time \cite{Bousso:2009dm,Bousso:2010id}, defined according to
\beq
t = -\frac{1}{3}\ln\!\big\{V_0\big[{\cal I}^+(p)\big]\big\} \,,
\eeq
where $V_0[{\cal I}^+(p)]$ is the volume on $\Sigma_0$ subtended by the subset of an initially uniform geodesic congruence orthogonal to $\Sigma_0$ that intersects the causal future ${\cal I}^+(p)$ of the point $p$ at which the lightcone time is being evaluated.  One can think of the geodesic congruence as a tool to project the asymptotic comoving volume of ${\cal I}^+(p)$ back onto $\Sigma_0$.  Thus, as one considers points $p$ progressively further to the future of $\Sigma_0$, this projection covers a progressively smaller volume on $\Sigma_0$, and the lightcone time increases according to negative one third of the logarithm of this (another, equivalent definition is given in \cite{Bousso:2010id}).

Rather than repeat the full analysis of the previous subsection, we simply refer to the results of \cite{Harlow:2011az}.  The crucial difference in comparison to scale-factor time is in the analogue of (\ref{Dt}), i.e.~the change in lightcone time between $\Sigma_0$ and a hypersurface of fixed FRW time $\ot$ in a bubble that nucleates on $\Sigma_0$.  In the case of lightcone time, this is \cite{Harlow:2011az}\footnote{The analysis of \cite{Harlow:2011az} does not keep track of all of the terms that we include here; for a more thorough analysis see \cite{Salem:2011mj}.  That paper actually studies CAH time, but in the coarse-grain limit CAH time is proportional to the exponent of lightcone time, allowing us to convert from the results of that paper.}
\beq
\Delta t = H_j\ot -\ln(2) + \xi
+\frac{2}{3}\ln\!\left(\frac{1}{2}+\frac{1}{2}e^{-\xi}\right) 
+\frac{1}{3}\ln\!\left[\frac{(H_i+H_je^{-\xi})^4}{(H_i+H_j)H_i^3}\right]\,.
\label{Dt2}
\eeq
Proceeding as in the previous section, the full consequence of this difference is to multiply (\ref{vol1}) by a factor of $H_i^3/H_j^3$, which corresponds to multiplying the second term in (\ref{r10}) by a factor of $H_j^3/H_i^3$.  Thus, we recover a rate equation of the form (\ref{r10}) if we work in terms of the rescaled volumes $\tilde{V}_i=H_i^3V_i$.  Likewise we define
\beq
\tilde{f}_i = \frac{\tilde{V}_i}{\sum_j \tilde{V}_j} = \frac{H_i^3V_i}{\sum_j H_j^3V_j}\,.
\label{tfi}
\eeq
Since in the coarse-grain approach the various $H_i$ are simply constants, this gives an equation of the form (\ref{r10b}), from which the solutions can be read off.

\section{Including bubble collisions}
\label{sec:coll}

The calculation of Section \ref{sec:rate1} neglects the effects of bubble collisions, which we now address.  

\begin{figure}[t!]
\begin{center}
\begin{tabular}{cc}
\includegraphics[width=0.45\textwidth]{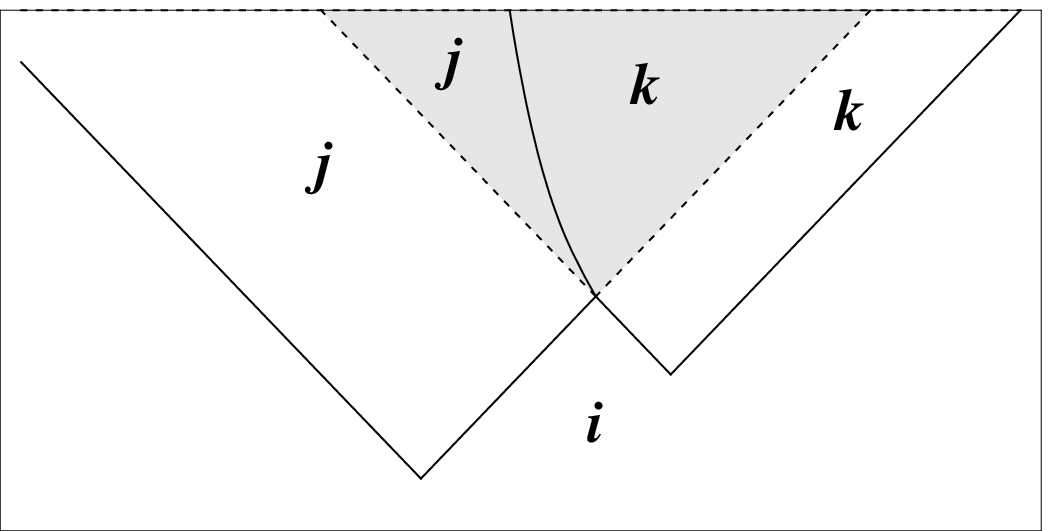} & 
\includegraphics[width=0.45\textwidth]{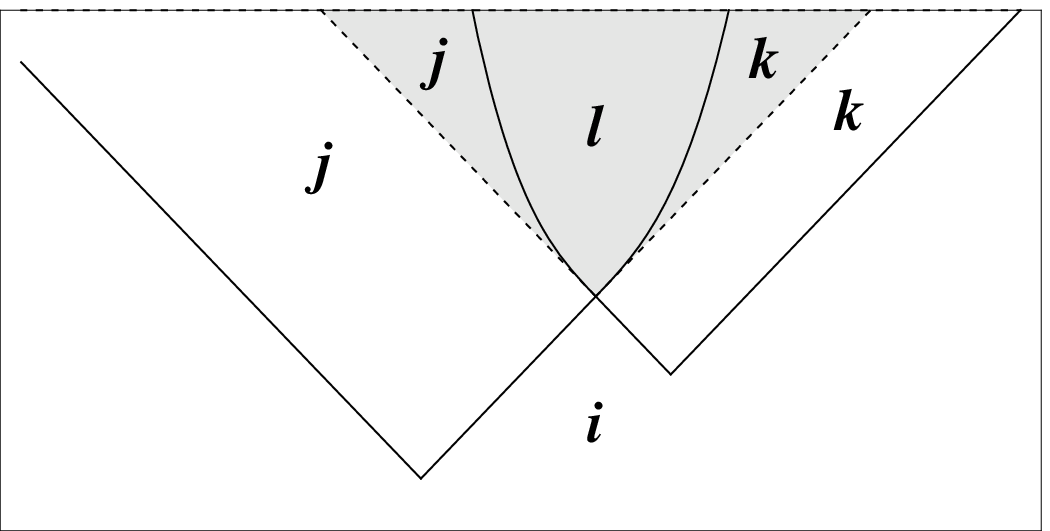}
\end{tabular}
\caption{\label{fig:coll}Left panel: toy spacetime diagram of a bubble collision producing a single domain wall.  The bubble walls are represented by solid curves following null rays until they collide, producing a domain wall.  Regions in vacua $i$, $j$, and $k$ are labeled; the causal future of the collision is bounded by dotted lines and shaded gray.  Right panel:  the same as in the left panel, except the collision produces two domain walls (a classical transition) enclosing vacuum $\ell$.}
\end{center}
\end{figure}

Consider the collision between one bubble of vacuum $j$ and another bubble of vacuum $k$, the two bubbles having nucleated in vacuum $i$ (see in Figure \ref{fig:coll}).  We assume that both $j$ and $k$ have smaller vacuum energies than $i$.  When the bubbles collide, either the bubble walls annihilate (if $j=k$), or one or two domain walls form after the collision.  In the case of one domain wall, the causal future of the collision contains both vacua $j$ and $k$, their relative volume fraction determined by the trajectory of the domain wall.  In the case of two domain walls, the domain walls contain some different vacuum $\ell\ne j,k$.  If the vacuum energy of $\ell$ is larger than the vacuum energies of $j$ and $k$, the domain walls accelerate toward each other. Nevertheless, if the vacuum energy of $\ell$ is less than the vacuum energy of $i$, the domain walls do not necessarily collide before future infinity.  This is called a classical transition to vacuum $\ell$ \cite{BlancoPillado:2009di,Easther:2009ft,Yang:2009wz,Johnson:2010bn,Deskins:2012tj}; in this case the causal future of the collision contains vacua $j$, $k$, and $\ell$, with the relative volume fractions determined by the trajectories of the domain walls.

We use $\lambda_{\ell kji}$ to denote volume fraction in vacuum $\ell$ in the causal future of a collision between bubbles of vacua $k$ and $j$ which nucleated in vacuum $i$.  As remarked above, for any given collision this volume fraction depends on the trajectories of domain walls.  These depend on the model-dependent interaction potential governing the tunneling fields, as well as on the collision-dependent placement of the colliding bubbles.  We are uninterested in these details, and instead simply take the $\lambda_{\ell kji}$ as given, taking the given quantities to average over the relative placement of colliding bubbles (further details are presented below).

\subsection{Volume corrections from bubble collisions in scale-factor time}
\label{ssec:insidedS}

Consider a dS bubble of vacuum $j$, which nucleates in some vacuum $i$ at $t=0$.  According to (\ref{vol1}), the physical volume in this bubble on the constant scale-factor hypersurface $\Sigma_{\Delta t}$ is  
\beq
V = \frac{4\pi}{3H_i^3}\,e^{3\Delta t} \,,
\label{futureV}
\eeq  
for large $\Delta t$.  However, some fraction of this volume is not actually in vacuum $j$, because it resides in the causal future of collisions between the bubble of vacuum $j$ and other bubbles.  To account for this, we first compute the total volume in the causal futures of these collisions, so that we can subtract this volume of $j$ from the rate equation.  We then discuss the volume of each vacuum that should be put back into the rate equation so as to reflect the vacuum composition of the causal futures of these collisions.  

Part of the calculation is laid out nicely in \cite{Dahlen:2008rd}, and we begin by translating the relevant result into our notation.  To do so, consider a 2-sphere of radius $\xi$ on a constant FRW time slice in the bubble geometry (\ref{bubblemetric}).  According to the symmetries of the collision, if the causal future of a collision intersects this 2-sphere, the intersection corresponds to a disk on the 2-sphere.  This disk subtends a certain solid angle, and we are interested in the total solid angle subtended by summing over the causal futures of all bubble collisions.  This is the calculation performed in \cite{Dahlen:2008rd}, and the resulting solid angle (in the limit of late FRW time slices) is  
\beq
4\pi\,f(\xi) = 4\pi\!\left[ 1-e^{-\Omega_{\rm tot}(\xi)/4\pi} \right]\,,
\label{sa}
\eeq
where
\beq
\Omega_{\rm tot}(\xi) =  4\pi\sum_k\kappa_{ki}\left\{ \frac{H_i^2}{H_j^2} 
+\ln\!\left[\frac{H_i^2}{H_j^2}+2\frac{H_i}{H_j}\cosh(\xi)+1\right]\right\} ,
\label{solida}
\eeq
where the sum runs over all vacua $k$ into which the vacuum $i$ can decay (with $\Lambda_k\leq\Lambda_i$).  We have used $\tan(T_{\rm co})=(H_i/H_j)\tanh(H_j\ot/2)\to H_i/H_j$ in converting from the notation of \cite{Dahlen:2008rd} in the context of the coarse-grain focus on the late-FRW-time limit in the bubble.

To obtain the physical volume on $\Sigma_{\Delta t}$ that intersects these causal futures, we integrate over 2-spheres of radii $\xi$, but using the solid angle (\ref{sa}) instead of the usual $4\pi$, and using the induced metric (\ref{induced}) to switch from a constant-$\ot$ to a constant-$t$ hypersurface.  This gives
\beq
\int_0^\infty d\xi\, 4\pi\sinh^2(\xi)\,H_i^{-3}\, e^{3\Delta t}
\left(\frac{1}{2}+\frac{1}{2}e^{-\xi}\right)^{\!\!-2}
e^{-3\xi}\, f(\xi) \equiv H_i^{-3}\,e^{3\Delta t}\, F\,,
\label{intermediate}
\eeq
where the second expression defines $F$, which is a dimensionless function of $R\equiv H_i/H_j$ and $\kappa_i\equiv \sum_k\kappa_{ki}$.  The integral in $F$ can be evaluated in terms of hypergeometric functions, but the result is not very illuminating.  Instead we focus on the situation where $R\gg 1$, and we assume $\kappa_i\ll 1$.  Then we can approximate 
\bea
F &=& \int_0^\infty d\xi\, 4\pi\sinh^2(\xi)
\left(\frac{1}{2}+\frac{1}{2}e^{-\xi}\right)^{\!\!-2}
e^{-3\xi}\,\left\{1-e^{-\kappa_iR^2}
\Big[R^2+2R\cosh(\xi)+1\Big]^{-\kappa_i}\right\}\,\,\,\,\,\,\,\, \\
&\approx& \int_0^{\ln(R)} d\xi\, 4\pi\sinh^2(\xi)
\left(\frac{1}{2}+\frac{1}{2}e^{-\xi}\right)^{\!\!-2}
e^{-3\xi}\left(1-e^{-\kappa_iR^2}R^{-2\kappa_i}\right) \nn\\*
& & +\, \int_{\ln(R)}^\infty d\xi\, 4\pi\,e^{-\xi}
\left[1-R^{-\kappa_i}e^{-\kappa_i(R^2+\xi)}\right] \\
&\approx& \frac{4\pi}{3}\frac{H_i^2}{H_j^2}\sum_k\kappa_{ki} \,,
\eea  
where we have kept only the leading-order terms in $H_i/H_j$ and $\kappa_{ij}$.  Combining with (\ref{intermediate}), we find that the physical volume in the causal future of these bubble collisions is  
\beq
\frac{4\pi}{3H_j^2H_i}\,e^{3\Delta t}\sum_k\kappa_{ki} \,.
\label{result2}
\eeq

Note that the volume subtracted from a bubble of vacuum $j$ due to a collision with a bubble of vacuum $k$ is not symmetric with respect to the indices $j$ and $k$.  In other words, when a bubble of vacuum $j$ collides with a bubble of vacuum $k$, the causal future of the collision intersects a different ``would-be'' volume on $\Sigma_{\Delta t}$ in the bubble of vacuum $j$, than the ``would-be'' volume on $\Sigma_{\Delta t}$ that it intersects in the bubble of vacuum $k$.  There is no reason these two volumes should have been the same, because neither of them represents the actual geometry in the causal future of the collision; instead they represent the volume that would have been there had these bubbles not collided, and this counterfactual volume is different between the two bubbles because the two bubbles have different Hubble rates.  

After we subtract the volume (\ref{result2}) from a bubble of vacuum $j$, we must specify how much volume of each type of vacuum to put back in its place.  We write the volume of vacuum $\ell$ on the hypersurface $\Sigma_{\Delta t}$ in the causal futures of the collisions between a bubble of vacuum $j$ and other bubbles of vacuum $k$, for all $k$, as
\beq
\frac{4\pi}{3H_\ell^2H_i}\,e^{3\Delta t}
\sum_k\kappa_{ki} \lambda_{\ell kji} \,,
\label{putback}
\eeq
where $i$ labels the vacuum in which the bubbles of vacua $j$ and $k$ nucleate.  Of course, all of the microphysics is contained in the $\lambda_{\ell kji}$, which are also understood to reflect the average volume fractions after considering the various possible relative placements of the colliding bubbles that contribute to each term in (\ref{putback}).  The other factors in (\ref{putback}) are chosen so that if collisions with bubbles of a given vacuum $k$ always produce a domain wall that runs along the causal future of the collision into the bubble of vacuum $k$---that is, from the perspective of the bubble of vacuum $j$, collisions with bubbles of vacuum $k$ have no effect on the interior of the bubble of vacuum $j$---then $\lambda_{\ell kji}=\delta_{\ell j}$ (this returns all of the volume that had been removed from the bubble of vacuum $j$, but none of the volume that had been removed from the bubbles of vacuum $k$).

The result (\ref{result2}) assumes that both $i$ and $j$ are dS vacua.  Since we ignore transitions out of AdS vacua, this assumption about $i$ is sufficient.  On the other hand, it is important to keep track of the case where the vacuum $j$ is AdS.  This is because it is possible for the collision between two AdS bubbles to create a classical transition to a dS vacuum.\footnote{The vacuum reached by a classical transition must have a smaller vacuum energy than the vacuum in which the colliding bubbles nucleate.  Therefore, two (AdS) bubbles that nucleate in an AdS vacuum cannot collide to produce a classical transition to a dS vacuum.  This is one reason why it remains unimportant to track bubble nucleations in AdS vacua.}  Guided by the results of Section \ref{sec:rate1} in which conservation of comoving volume implied that after stripping away the volume expansion factors the same expressions could be used for transitions to dS and AdS vacua, we assume that (\ref{result2}) can be used to describe the volume in the causal future of bubble collisions for both dS and AdS vacua $j$, modulo the volume expansion factor $e^{3\Delta t}$.  Likewise, we use (\ref{putback}) for both dS and AdS vacua $\ell$ (and $j$ and $k$).  Although it is possible that using (\ref{result2}) for AdS vacua $j$ introduces an error, technically this error can be removed by an appropriate choice for the factors $\lambda_{\ell kji}$ in (\ref{putback}).  Our analysis is unconcerned with such details; in our analysis it is only important whether $\lambda_{\ell kji}$ is precisely zero or not.

\subsection{Modifying the rate equation in scale factor time}

We can now compute the change in volume in vacuum $i$, $\Delta V_i$, after a scale-factor time interval $\Delta t$, due to bubble collisions.  We first consider the loss of volume in bubbles of vacuum type $i$ due to collisions with other bubbles.  This is
\beq
\Delta V_i = -\sum_{j,k}\frac{\Gamma_{ij}}{H_j}\,V_j\,\Delta t\,
\frac{4\pi}{3H_i^2H_j}\,\kappa_{kj} = -\sum_{j,k}\kappa_{ij}\kappa_{kj} 
\frac{H_j^2}{H_i^2}\, V_j\,\Delta t \,.
\eeq
The first three terms in the first expression give the number of bubbles of vacuum $i$ that nucleate in vacuum $j$ in the time $\Delta t$, according to (\ref{dnumber}).  The other terms in this expression give the volume in each such bubble that is in the causal future of collisions with bubbles of vacuum $k$, given by (\ref{result2}), but where we have removed the expansion factor $e^{3\Delta t}$, which in the coarse-grain analysis is accounted for by the factor of $3V_i$ in (\ref{r10}). 

Next we consider the change in volume in vacuum $i$, due to the physical volume in vacuum $i$ in the causal future of bubble collisions.  This can be written
\beq
\Delta V_i = \sum_{j,k,\ell}\frac{\Gamma_{\ell j}}{H_j}\,V_j\,\Delta t\,
\frac{4\pi}{3H_i^2H_j}\,\kappa_{kj}\,\lambda_{ik\ell j} 
= \sum_{j,k,\ell} \lambda_{ik\ell j}\,\kappa_{kj} \kappa_{\ell j}
\frac{H_j^2}{H_i^2}\,V_j\,\Delta t \,.
\label{vstep}
\eeq
The first three terms in the first expression give the number of bubbles of type $\ell$ that nucleate in vacuum $j$ in the time $\Delta t$, according to (\ref{dnumber}).  The other terms in this expression give the volume in vacuum $i$ in the causal futures of the collisions between each such bubble of vacuum $\ell$ and bubbles of vacuum $k$, given by (\ref{putback}), but where we have removed the expansion factor $e^{3\Delta t}$, which in the coarse-grain analysis is accounted for by the factor of $3V_i$ in (\ref{r10}).  

Putting everything together and taking the infinitesimal limit, we obtain
\beq
\frac{dV_i}{dt} = 3V_i + \sum_j\kappa_{ij}V_j-\sum_j\kappa_{ji}V_i 
+\sum_{j,k,\ell} \lambda_{ik\ell j}\kappa_{kj}\kappa_{\ell j}\frac{H_j^2}{H_i^2}V_j
-\sum_{j,k}\kappa_{ij}\kappa_{kj}\frac{H_j^2}{H_i^2}V_j \,.
\label{rate2}
\eeq
In terms of the volume fractions $f_i$ defined in (\ref{fidef}), we have    
\beq
\frac{df_i}{dt} = M_{ij}f_j \,, 
\label{r10c}
\eeq
where now 
\beq  
M_{ij} \equiv \kappa_{ij} - \delta_{ij}\sum_k\kappa_{ki}
+\sum_{k,\ell}\gamma_{ik\ell j}\kappa_{kj}\kappa_{\ell j}\,,
\label{mdef2}
\eeq
where for later convenience we have defined 
\beq
\gamma_{ik\ell j} \equiv \frac{H_j^2}{H_i^2}\left(\lambda_{ik\ell j} - \delta_{i\ell}\right) \,.
\label{gdef}
\eeq
Note that conservation of comoving volume implies that the sum over rows in any column of $M_{ij}$ must be zero.  This implies a useful constraint on the $\gamma_{ik\ell j}$, namely
\bea
\sum_{i,k,\ell} \gamma_{ik\ell j} = 0 \,. 
\label{gconstraint}
\eea

\subsection{Lightcone time}

The previous two subsections worked in terms of scale-factor time.  However, the only place where the actual definition of the global time parameter enters is in the induced metric used in (\ref{intermediate}) (and results, such as (\ref{futureV}), taken from Section \ref{sec:rate1}).  Meanwhile, in Section \ref{ssec:lct} we found that the only effect of using lightcone time instead of scale-factor time corresponded to multiplying the second term in (\ref{r10}) by $H_j^3/H_i^3$.  The same applies here with respect to the second term in (\ref{rate2}), and because of the modification to (\ref{intermediate}) we multiply the fourth and fifth terms in (\ref{rate2}) by $H_j^3/H_i^3$ as well.  It is easily checked that if we define $\tilde{f}_i$ and $\tilde{V}_i$ as in (\ref{tfi}) and above it, we obtain
\beq
\frac{d\tilde{f}_i}{dt} = M_{ij}\tilde{f}_j \,,
\label{rate2lct}
\eeq
with $M_{ij}$ given by (\ref{mdef2}).  Therefore, given a solution in terms of scale-factor time, we can again read off the solution in terms of lightcone time.

\section{Phenomenology}
\label{sec:phenomenology}

To develop intuition for the phenomenology of the rate equation, we study a simple toy model of the landscape \cite{Garriga:2005av,Linde:2006nw,DeSimone:2008if}.  The model can be represented by the diagram
\beq
\begin{array}{ccccc}
1 & \leftarrow\hspace{-8pt}\rightarrow & 2 & \leftarrow\hspace{-8pt}\rightarrow & 3  \\
\vspace{-18pt} & & & &  \\
\downarrow & & \downarrow & & \downarrow \\
\vspace{-16pt} & & & & \\
4 & & 5 & & 6 \\
\end{array} \,\,\,\,\,.
\label{eq:modelA}
\eeq
The numbers label different vacua, while the arrows indicate the direct transitions that are allowed among the vacua.  (We use the phrase ``direct transition'' to designate semiclassical bubble formation via quantum tunneling through one potential barrier, which we take to be the dominant form of vacuum transition, aside from perhaps classical transitions, which are discussed below.)  For concreteness, we assume the vacuum energies of this model obey 
\beq
\Lambda_4,\,\Lambda_5,\,\Lambda_6 < 0 < \Lambda_1,\,\Lambda_3 < \Lambda_2 \,.
\label{hierarchyA}
\eeq  
That is, the labels ``1,'' ``2,'' and ``3'' designate dS vacua, while the other labels designate AdS vacua.  We also assume that ``1'' corresponds to the dS vacuum with the smallest decay rate.

\subsection{Toy model without bubble collisions}
\label{ssec:nocollisionsA}

We first ignore bubble collisions.  Then this model is similar to models studied elsewhere in the literature \cite{Garriga:2005av,Linde:2006nw,DeSimone:2008if}.  Seeking the late-time attractor solution---that is, the dominant eigenvector---we insert the ansatz $f_i,\tilde{f}_i = s_i e^{-qt}$ into the rate equation, obtaining
\bea
-q s_1 &=& \kappa_{12} s_2 - \kappa_1 s_1 \\
-q s_2 &=& \kappa_{21} s_1 + \kappa_{23} s_3 - \kappa_2 s_2 \\
-q s_3 &=& \kappa_{32} s_2 - \kappa_3 s_3 \,,
\eea  
where we focus on the volume fractions of the dS vacua, and $\kappa_i\equiv\sum_j\kappa_{ji}$.  This system of equations can be solved algebraically, but the solution is complicated and not very enlightening.  Since $\kappa_{21}$ and $\kappa_{23}$ are transition rates from lower to higher vacuum energy, they are expected to be exponentially suppressed relative to the other transition rates.  We therefore solve the above system of equations by expanding in $\kappa_{21}$ and $\kappa_{23}$, treating them as similar in order for the purpose of the expansion.  

At zeroth order the only nonzero dS component of $s_i$ is $s_1$, which we can set to unity.  The corresponding value of $q$ is 
\beq
q = \kappa_{41}+\kappa_{51} = \kappa_1 \,,  
\eeq  
where the second equality implicitly evaluates $\kappa_1$ at zeroth order, a notational shorthand that we also employ below to help simplify expressions.  The other dS components of $s_i$ become relevant at first order, and are given by
\bea
s_2 = \frac{\kappa_{21}}{\kappa_2-q} \,,\qquad
s_3 = \frac{\kappa_{21}\kappa_{32}}{(\kappa_2-q)(\kappa_3-q)} \,.
\label{domA}
\eea
The zeroth-order solutions for $s_1$ and $q$ receive first-order corrections, but these are subdominant and so we ignore them.  We see that the dS components of $s_i$ are dominated by ``1''---the dS vacuum with the smallest decay rate---which is accordingly called the dominant vacuum (this assumes that the decay rates are not tuned so as to make $q$ very near to $\kappa_2$ or $\kappa_3$, in which case there could be a set of degenerate dominant vacua \cite{DeSimone:2008if}).  Intuitively, the dominant vacuum dominates the asymptotic volume of spacetime due to its small decay rate, and the volume fractions of all of the other vacua reflect in part the relative likelihood of transitioning to them from the dominant vacuum.  In particular, $s_2$ carries a factor of $\kappa_{21}$, reflecting the transition from ``1'' to ``2,'' while $s_3$ carries the factor $\kappa_{21}\kappa_{32}$, reflecting transitions first from ``1'' to ``2,'' then from ``2'' to ``3.''  These factors also contain total decay rates in the denominator, so they only correspond to suppression factors when a transition rate in the numerator is relatively suppressed.  This applies to $\kappa_{21}$ in the first-order terms in (\ref{domA}), and it applies to contributions to the $s_i$ that arise due to transitions through ``$3$,'' which are suppressed relative to the above contributions.

\subsection{Effects of bubble collisions}
\label{ssec:collisions}

The above discussion summarizes previous work.  Our goal is to understand the implications of allowing for bubble collisions.  Collisions that result in a single domain wall, and therefore merely shift the volume fractions of the vacua involved in the collision by order-unity factors, are not consequential in the context of the qualitative dynamics described above.  However, the possibility of classical transitions might change the above conclusions in more dramatic ways.  To explore this, we first assume that bubbles of vacua ``4'' can collide and produce a classical transition to vacuum  ``3.'' We denote this with the diagram 
\beq
4)(4 \to 3 \,.
\label{collisionA}
\eeq
Note that this transition requires that $\Lambda_3<\Lambda_1$ \cite{Easther:2009ft,Johnson:2010bn}.  Although the landscape (\ref{eq:modelA}) per\-mits other bubble collisions, for the moment we assume that these do not produce classical transitions so that we can ignore them.  Then the full set of nonzero $\gamma_{ijk\ell}$ can be taken to be
\bea
\gamma_{4441} = -\frac{H_1^2}{H_3^2} \,,\qquad \gamma_{3441} = \frac{H_1^2}{H_3^2} \,.
\label{alphadef}
\eea
The first term accounts for the subtraction of volume in the causal future of the collisions that would be in vacuum ``4'' but for the collision, while the second term accounts for the restoration of this volume but in vacuum ``3,'' a consequence of the classical transition.  In a realistic model we expect some of the causal future of the collision to be in ``4,'' which means that $\gamma_{4441}$ should be less negative and $\gamma_{3441}$ should be smaller.  However, these effects change our results by factors that are negligible next to the exponential staggering of decay rates, and so for simplicity we ignore them.

Including these collisions and their effects, the rate equation gives   
\bea
-q s_1 &=& \kappa_{12} s_2 - \kappa_1 s_1 \\
-q s_2 &=& \kappa_{21} s_1 + \kappa_{23} s_3 - \kappa_2 s_2 \\
-q s_3 &=& \kappa_{32} s_2 - \kappa_3 s_3 +\gamma_{3441}\kappa_{41}^2 s_1\,,
\eea
where again we focus on the dS components of $s_i$.  As before, the solution is more transparent if we expand it in terms of $\kappa_{21}$ and $\kappa_{23}$.  At zeroth order, we find
\beq
s_1 = \left(1+\frac{\gamma_{3441}\kappa_{41}^2}{\kappa_3-q}\right)^{\!-1} ,\qquad
s_3 = \frac{\gamma_{3441}\kappa_{41}^2}{\kappa_3-q} \,,
\eeq
where we have normalized the zeroth-order solution so that $\sum_is_i=1$, where the sum runs over dS vacua $i$, and $q$ is unchanged from before:  $q=\kappa_1$.  The leading-order contribution to $s_2$ still appears at first order in the expansion,
\bea
s_2 = \frac{\kappa_{21}s_1}{\kappa_2-q}+\frac{\kappa_{23}s_3}{\kappa_2-q} \,,
\label{domA2}
\eea
where $s_1$ and $s_3$ refer to the zeroth-order quantities.  Evidently, the hierarchical structure of the dominant eigenvector is qualitatively changed relative to before.  In particular, while the dominant vacuum ``1'' still dominates the dS components of $s_i$, the component $s_3$ now contains a product of two downward transition rates from ``1,'' as opposed to an upward transition rate.  Note that since ``1'' is by definition the dS state with the smallest decay rate, we still expect $s_3$ to be exponentially suppressed relative to $s_1$.  Nevertheless, the size of the suppression is dramatically reduced.  Meanwhile, the volume fraction of ``2'' is still suppressed by an upward transition rate, but this can come from ``1'' or from ``3.'' 

Although the results are changed from when we ignored bubble collisions, our intuition for the solution is maintained.  In particular, the dS vacuum with the smallest decay rate dominates the asymptotic volume fraction, and the volume fractions of the other dS vacua reflect the relative likelihood of transitioning to them from the dominant vacuum.  In the case of ``3,'' this likelihood is enhanced by the possibility for classical transitions.  This in turn enhances the volume fraction of ``2,'' insofar as it can be reached by transitions from ``3.''   

The classical transition represented by (\ref{collisionA}) is not the only possibility in the landscape (\ref{eq:modelA}).  Another possibility occurs when bubbles collide in vacuum ``2.''  We now focus on the possibility that collisions between bubbles of vacua ``5'' can result in a classical transition to ``3,'' represented by the diagram
\beq
5)(5 \to 3 \,,
\eeq
and ignore all other possible collisions.  Then the full set of nonzero $\gamma_{ijk\ell}$ can be taken to be
\bea
\gamma_{5552} = -\frac{H_2^2}{H_3^2} \,,\qquad
\gamma_{3552} = \frac{H_2^2}{H_3^2} \,,
\label{betade}
\eea
where the discussion below (\ref{alphadef}) applies here as well.  The resulting rate equation gives  
\bea
-q s_1 &=& \kappa_{12} s_2 - \kappa_1 s_1 \\
-q s_2 &=& \kappa_{21} s_1 + \kappa_{23} s_3 - \kappa_2 s_2 \\
-q s_3 &=& \kappa_{32} s_2 - \kappa_3 s_3 +\gamma_{3552}\kappa_{52}^2 s_2\,.
\eea
We again expand in terms of $\kappa_{21}$ and $\kappa_{23}$.  The zeroth order dS components are the same as without classical transitions; namely the only nonzero component is $s_1$, which we can set to unity.  The other dS components of $s_i$ become relevant at first order, and are given by    
\beq
s_2 = \frac{\kappa_{21}}{\kappa_2-q}\,,\qquad
s_3 = \frac{(\kappa_{32}+\gamma_{3552}\kappa_{52}^2)\,\kappa_{21}}{(\kappa_2-q)(\kappa_3-q)} \,.
\label{domA3}
\eeq
Note that $s_2$ and $s_3$ are still suppressed by an upward transition rate out of ``1.''  Therefore, the effects of classical transitions do not change the qualitative expectations for the dominant eigenvector described above.  On the other hand, it is possible for the component $s_3$ to be enhanced relative to before, depending on the relative size of $\gamma_{3552}\kappa_{52}^2$ and $\kappa_{32}$.

\subsection{Generalization of the toy model}
\label{ssec:generalize}

The toy landscape model (\ref{eq:modelA}) is simple, but the intuition developed above extends to more general landscapes.  Indeed, it is possible to compute the components of the dominant eigenvector $s_i$ for any number of vacua and for any set of transitions among them, by expanding in the off-diagonal elements of the transition matrix $M_{ij}$ \cite{Olum:2007yk,Bousso:2012es}.  As before, we assume the dS vacuum with the smallest decay rate is unique and denote it as ``1.'' The (unnormalized) dS components of $s_i$ are then, to leading order,
\beq
s_i = \sum_{\{p_a\}}\, \frac{\kappa_{ip_1}}{\kappa_i-q} 
\times\ldots\times\frac{\kappa_{p_n1}}{\kappa_{p_n}\!-q}
\approx \sum_{\{p_a\}}\, \frac{q}{\kappa_i}\,\frac{\kappa_{ip_1}}{\kappa_{p_1}}
\times\ldots\times\frac{\kappa_{p_n1}}{q} \,,
\label{safull}
\eeq
where $q$ is the decay rate of ``1'' and the sum covers the sequences of transitions that connect ``1'' to $i$ using the fewest number of upward transitions.  In the second expression we have exploited the fact that $q$ is the smallest decay rate to write $\kappa_i-q\approx \kappa_i$, which allows for the terms to be rearranged so as to express factors as branching ratios.  

We expect the dS vacuum with the smallest decay rate to also possess a small vacuum energy (for some discussion of this matter see \cite{DeSimone:2008if}).  This ensures that transitions from ``1'' to other dS states are either transitions to larger vacuum energy or transitions that involve a very small change in vacuum energy compared to the expected size of the barrier.  The rates associated with such transitions are exponentially suppressed relative to the rates associated with transitions to AdS vacua with much larger magnitude vacuum energies \cite{Coleman:1980aw}.  Therefore, we expect the rightmost term in (\ref{safull}) to be exponentially small, while all of the other factors are less than one by definition.  This maintains the notion of ``1'' as the state that dominates the volume fractions, i.e.~the dominant vacuum.  

It is straightforward to generalize the above result to include the effects of bubble collisions.  In particular, the effect of bubble collisions is simply to add the (small) quantity $\sum_{k,\ell}\gamma_{ik\ell j}\kappa_{kj}\kappa_{\ell j}$ to the quantity $\kappa_{ij}$ corresponding to each off-diagonal perturbative element of the transition matrix $M_{ij}$.  We can then read off the results from the analyses of \cite{Olum:2007yk,Bousso:2012es}: 
\bea
s_i &=& \sum_{\{p_a\}}\,\frac{\kappa_{ip_1}\!+\sum_{j,k}\gamma_{ijkp_1}\kappa_{jp_1}\kappa_{kp_1}}{\kappa_i-q} 
\times\ldots\times\frac{\kappa_{p_n1}+\sum_{j,k}\gamma_{p_njk1}\kappa_{j1}\kappa_{k1}}{\kappa_{p_n}\!-q} \\
&\approx& \sum_{\{p_a\}}\,\, \frac{q}{\kappa_i}\,
\frac{\kappa_{ip_1}\!+\sum_{j,k}\gamma_{ijkp_1}\kappa_{jp_1}\kappa_{kp_1}}{\kappa_{p_1}} 
\times\ldots\times\frac{\kappa_{p_n1}+\sum_{j,k}\gamma_{p_njk1}\kappa_{j1}\kappa_{k1}}{q} \,. \qquad
\label{safull2}
\eea
Of course, it is not necessary for both terms in a given factor to be nonzero, and in fact the consequences of bubble collisions are most significant when the first term in at least one of these factors is zero, as with the classical transitions described above.

Note that the dS vacuum with the smallest decay rate, vacuum ``1,'' still dominates the volume fractions.  To elaborate, we repeat the argument given above for the case where bubble collisions are ignored, which holds unless the first term in the numerator of the rightmost factor of (\ref{safull2}) is zero.  This occurs when a dS vacuum $i$ can be reached via classical transitions caused by bubble collisions in ``1.''  In this case, each contribution to (\ref{safull2}) contains a factor of the form $\kappa_{i1}/q$, which is necessarily less than one, in addition to a factor of the form $\gamma_{ijk1}\kappa_{k1}$, which we expect to be much less than one.  The latter expectation arises because although any $\gamma_{ijk1}$ involves a ratio of Hubble rates (squared), which in principle could be very large, the factor $\kappa_{k1}$ is expected to be doubly exponentially small, as it is less than the decay rate per Hubble volume of the longest-lived dS state.    

Evidently, classical transitions can modify the detailed hierarchical structure among the components of the dominant eigenvector.  For instance, a given ratio $s_i/s_j$ could be exponentially large or small depending on whether one ignores classical transitions or not.  Moreover, the existence of classical transitions can modify the distribution of components of $s_i$ as a function of their size.  To illustrate these conclusions, we organize the vacua in the landscape according to the number $\Nup$ of upward transition rates that appear in the numerators of the factors in the corresponding components of $s_i$.  For instance, a vacuum $i$ for which the sequences of transitions $\{p_a\}$ in the summation of (\ref{safull2}) contain three upward transitions has $\Nup=3$.  Accounting for classical transitions can allow for sequences of transitions that reduce $\Nup$ relative to otherwise, for a given vacuum in the landscape.  Since upward transition rates are exponentially suppressed relative to downward transition rates, this exponentially increases the volume fractions of certain vacua (relative to not including classical transitions), and it modifies the distribution of vacua as a function of volume fraction.

This is relevant to the staggering issue, which concerns the competition between anthropic selection for a small (magnitude) cosmological constant $\Lambda$ and cosmological selection for a large volume fraction when attempting to explain the observed value of $\Lambda$ \cite{SchwartzPerlov:2006hi,SchwartzPerlov:2006hz,Olum:2007yk,SchwartzPerlov:2008he}.  To elaborate, imagine organizing the various vacua in the landscape into bins according to the size of $\Lambda$, and consider the total volume fraction represented by the vacua in each bin.  The landscape explanation of the observed value of $\Lambda$ assumes that for a bin size $\Delta\Lambda$ less than the observed value of $\Lambda$, the total volume fraction in each bin is roughly independent of $\Lambda$ for small $|\Lambda|$.  The validity of this assumption depends on the number of vacua in the landscape as well as the distribution of these vacua as a function of volume fraction.  If the number of vacua is too small, then the number of vacua in each bin will be small and their total volume fraction will vary wildly from bin to bin.  If the distribution of vacua as a function of volume fraction falls too steeply, then the total volume fraction in each bin will be dominated by the volume fractions of a few vacua, and again the total volume fraction will vary wildly from bin to bin.  In either case, the anthropic suppression associated with a set of vacua with $\Lambda$ many orders of magnitude larger than the value we observe could be compensated by a much larger volume fraction occupied by these vacua.  The quantitative analysis is model dependent and further discussion of this issue is beyond the scope of this paper.

\subsection{Boltzmann brains}
\label{ssec:BBs}

The volume fractions represented by $s_i$ are also relevant to the Boltzmann-brain issue, which concerns the likelihood for an observer to arise after reheating from a relatively low-entropy inflationary state, as opposed to after a quantum fluctuation from a relatively high-entropy vacuum-energy-dominated state \cite{Dyson:2002pf,Albrecht:2004ke,Page:2006dt,Bousso:2006xc,Bousso:2011aa}.  The former observers are called normal observers and the latter observers are called Boltzmann brains.  To discuss this issue, it is helpful to first summarize the case where bubble collisions are ignored.  With respect to the scale-factor cutoff measure, the ratio of Boltzmann brains to normal observers can be approximated by \cite{DeSimone:2008if}
\beq
\frac{{\cal N}^{\mbox{\tiny BB}}}{{\cal N}^{\mbox{\tiny NO}}} 
\sim \frac{\sum_i \kappa^{\mbox{\tiny BB}}_is_i^{\phantom{i}}}
{\sum_{i,j} n^{\mbox{\tiny NO}}_{ij} \kappa_{ij}^{\phantom{i}}s_j^{\phantom{i}}} \,,
\label{BBtoNO}
\eeq 
where $\kappa^{\mbox{\tiny BB}}_i$ is the Boltzmann-brain formation rate per unit Hubble volume in vacuum $i$, $n^{\mbox{\tiny NO}}_{ij}$ is the peak number of normal observers per unit Hubble volume in a bubble of vacuum $i$ that nucleates in vacuum $j$, and the sums are understood to run over only dS vacua.  

To analyze (\ref{BBtoNO}), first note that Boltzmann-brain formation rates are generically doubly exponentially small.  For example, simply demanding that a Boltzmann brain possess at least the information content of a human brain gives the upper bound $\kappa^{\mbox{\tiny BB}}_i \lesssim e^{-10^{16}}\!$ \cite{DeSimone:2008if}.  Second, note that any transition rate out of a dS vacuum $i$ must exceed the recurrence rate in $i$, i.e. $\kappa_{ij} > e^{-3\pi/G\Lambda_j}$.  Therefore, assuming that there exists a sequence of transitions that connects the dominant vacuum ``1'' to a given vacuum $i$ such that no vacuum between ``1'' and $i$ has $G\Lambda\lesssim -1/\ln(\kappa^{\mbox{\tiny BB}}_i)$, we can use (\ref{safull}) to write \cite{Bousso:2008hz}  
\beq
\sum_i \kappa^{\mbox{\tiny BB}}_is_i^{\phantom{i}} 
\sim \max\{\kappa^{\mbox{\tiny BB}}_is_i^{\phantom{i}}\}
\sim \max\!\left\{\frac{\kappa^{\mbox{\tiny BB}}_1}{q},\,
e^{-3\pi/G\Lambda_1}\frac{\kappa^{\mbox{\tiny BB}}_i}{\kappa_i}\right\} ,
\label{num1}
\eeq
where we use the arithmetic of double exponentials (whereby if $x$ is a double exponential and $|\ln(y)|<|\ln(x)|$, then $xy\sim x/y\sim x$) to ignore the various factors of $\kappa_{jk}/\kappa_k$ in $s_i$ next to $\kappa^{\mbox{\tiny BB}}_i$, applying the principle of detailed balance to write $\kappa^{\mbox{\tiny BB}}_i\kappa_{j1}=\kappa^{\mbox{\tiny BB}}_i\,e^{3\pi/G\Lambda_j-3\pi/G\Lambda_1}\kappa_{1j}\sim e^{-3\pi/G\Lambda_1}\kappa^{\mbox{\tiny BB}}_i$.  Finally, assuming that there exists a sequence of transitions that connects ``1'' to a vacuum $i$ with $|\ln(n^{\mbox{\tiny NO}}_{ij})| < |\ln(\max\{\kappa^{\mbox{\tiny BB}}_k\})|$, such that no vacuum between ``1'' and $i$ has $G\Lambda\lesssim -1/\ln(\kappa^{\mbox{\tiny BB}}_i)$, we can apply the same techniques to the denominator of (\ref{BBtoNO}) to obtain
\beq
\frac{{\cal N}^{\mbox{\tiny BB}}}{{\cal N}^{\mbox{\tiny NO}}} 
\sim \max\!\left\{ e^{3\pi/G\Lambda_1}\frac{\kappa^{\mbox{\tiny BB}}_1}{q},\,
\frac{\kappa^{\mbox{\tiny BB}}_i}{\kappa_i}\right\} .
\label{BBresult}
\eeq
Note that $e^{3\pi/G\Lambda_1}$ sets the scale for the recurrence time in the dominant vacuum.  Therefore, if it is possible for Boltzmann brains to form in the dominant vacuum, the first term in brackets is much greater than one and Boltzmann brains dominate over normal observers.  Consequently, we assume that the dynamics of the dominant vacuum are insufficient to support Boltzmann brains---a plausible assumption considering the complex dynamics that underlies our existence.  Then, normal observers dominate if we furthermore assume that in each vacuum the decay rate is larger than the Boltzmann-brain formation rate.        

It is straightforward to repeat this analysis including the effects of bubble collisions.  The denominator of (\ref{BBtoNO}) essentially expresses the bubble formation rate, and generalizes to
\beq
\sum_{i,j} n^{\mbox{\tiny NO}}_{ij} \kappa_{ij}^{\phantom{i}}s_j^{\phantom{i}} \to 
\sum_{i,j} n^{\mbox{\tiny NO}}_{ij} \kappa_{ij}^{\phantom{i}}s_j^{\phantom{i}} +
\sum_{i,j,k,\ell} (H_j/H_i)^2 n^{\mbox{\tiny NO}}_{ik\ell j}\lambda_{ik\ell j}^{\phantom{i}}
\kappa_{kj}^{\phantom{i}}\kappa_{\ell j}^{\phantom{i}} s_j^{\phantom{i}} \,,
\eeq
where $n^{\mbox{\tiny NO}}_{ik\ell j}$ is defined in analogy to $n^{\mbox{\tiny NO}}_{ij}$ and the sums over $k$ and $\ell$ are understood to include AdS vacua.  The subsequent analysis is unchanged except for the possibility that Boltzmann brains form in a dS vacuum that can be reached via classical transitions caused by collisions between AdS bubbles in the dominant vacuum.  (This affects the analysis because we cannot use detailed balance to rewrite $\kappa_{j1}$ in terms of $\kappa_{1j}$ for an AdS vacuum $j$.)  Accounting for this additional possibility, we find     
\beq
\frac{{\cal N}^{\mbox{\tiny BB}}}{{\cal N}^{\mbox{\tiny NO}}} 
\sim \max\!\left\{ e^{3\pi/G\Lambda_1}\frac{\kappa^{\mbox{\tiny BB}}_1}{q},\,
\frac{\kappa^{\mbox{\tiny BB}}_i}{\kappa_i},\,
e^{3\pi/G\Lambda_1}\frac{\kappa^{\mbox{\tiny BB}}_\ell}
{\kappa_\ell}\gamma_{\ell jk1}\kappa_{j1}\kappa_{k1}\right\} ,
\label{BBresult2}
\eeq
where we have assumed that any vacua reached by classical transitions from bubble collisions in the dominant vacuum do not produce normal observers (as would be the case if for example these transitions do not establish a period of slow-roll inflation followed by reheating).  Note that if normal observers are produced in these vacua, then we simply obtain (\ref{BBresult}) again, since factors of the form $e^{3\pi/G\Lambda_1}\gamma_{\ell jk1}\kappa_{j1}\kappa_{k1}$ cancel between the numerator and denominator of (\ref{BBtoNO}) (according to the algebra of double exponentials).
  
As with (\ref{BBresult}), to avoid Boltzmann-brain domination we must assume that in each vacuum the decay rate is larger than the Boltzmann-brain formation rate, and that the dominant vacuum does not support Boltzmann brains.  Given these assumptions, the last term in the brackets does not exceed $e^{3\pi/G\Lambda_1}\kappa^{\mbox{\tiny BB}}_i$ for any of the relevant vacua $i$.  Therefore, we still avoid Boltzmann-brain domination unless $G\Lambda_1 < -1/\ln(\kappa^{\mbox{\tiny BB}}_i) < 10^{-16}$.  While such small values of $\Lambda$ are presumably very uncommon in a Planck-scale landscape, the dominant vacuum ``1'' is specially selected for the smallness of its decay rate, and it seems plausible that this might also select for an extremely small value of $\Lambda$.  Accepting this as a possibility, the next question is how plausible is it that ``1'' can set up a classical transition to a dS vacuum $\ell$ that supports Boltzmann brains, bearing in mind that vacua with sufficiently complex degrees of freedom and interactions are presumably rare, and the classical transition only occurs if $\Lambda_\ell < \Lambda_1$.          

To explore this question, we consider a large landscape of $N$ vacua to arise in a roughly $\log(N)$-dimensional configuration space \cite{Bousso:2000xa,Douglas:2003um,ArkaniHamed:2005yv}.  Thus, we assume that each vacuum can directly transition to roughly $\log(N)$ nearby states.  Classical transitions are permitted based on special relationships among local minima in the vacuum configuration space, such that each unique pair of vacua involved in a collision can classically transition to at most one unique state \cite{Easther:2009ft,Yang:2009wz,Deskins:2012tj}.  Hence, there are at most roughly $\log^2(N)$ distinct states that can be reached by collisions consequent the direct transitions mentioned above.  In the context of string theory, it is commonly argued that $\log(N)\sim {\cal O}(1000)$ or so \cite{Bousso:2000xa,Douglas:2003um}.  Based on these considerations, it seems very unlikely that the dominant vacuum could set up a classical transition to a vacuum $i$ with $0 < G\Lambda_i < -1/\ln(\kappa^{\mbox{\tiny BB}}_i)< 10^{-16}$, let alone such a vacuum that can also support Boltzmann brains.  

On the other hand, it is not evident that we should only consider direct transitions out of the dominant vacuum.  Naively, if ``1'' transitions to $i$ at the rate $\kappa_{i1}$ and if $i$ transitions to $j$ at the rate $\kappa_{ji}$, then ``1'' transitions to $j$ at roughly the rate $\kappa_{ji}\kappa_{i1}$.  Although $\kappa_{ji}\kappa_{i1}$ is exponentially suppressed relative to $\kappa_{i1}$, the factor $e^{3\pi/G\Lambda_1}$ in the last term in the brackets of (\ref{BBresult2}) is doubly exponentially large, and it seems possible that it could take an exponential number of factors of transition rates to cancel it.  If so, classical transitions set up by the dominant vacuum could access a significant fraction of the landscape, presumably including many dS vacua capable of supporting Boltzmann brains.  On the other hand, if these classical transitions also access a dS vacuum capable of supporting normal observers, then granting the assumption that $\kappa^{\mbox{\tiny BB}}_i < \kappa_i^{\phantom{i}}$ for all dS vacua $i$, these normal observers will dominate over all of the Boltzmann brains reached in this way.  Recently, it has been argued that the number of vacua capable of supporting Boltzmann brains is ``only'' exponentially (as opposed to doubly exponentially) larger than the number of vacua capable of forming normal observers \cite{Yang:2012jf}.  (Specifically, \cite{Yang:2012jf} argues that the number of bubbles that feature a significant amount of slow-roll inflation is ``only'' exponentially suppressed relative to the number that do not.)  If so, it would seem to take a fine-tuning for the dominant vacuum to set up classical transitions that reach enough vacua to include one supporting Boltzmann brains, but not enough so as to also include one that forms normal observers.

\subsection{Probability to observe a classical transition}
\label{ssec:past}

So far our discussion has focused on global volume fractions.  On the other hand, if the initial conditions for the phase of slow-roll inflation believed to describe our past were established by a classical transition---as opposed to by the formation of a single bubble---then this could have consequences for observational cosmology, at least if slow-roll inflation did not last too long.  Indeed, the collision geometry retains an SO(2,1) symmetry \cite{Chang:2007eq}, and so the phenomenological consequences should resemble those of the ``anisotropic bubble nucleation'' scenario studied in \cite{BlancoPillado:2010uw,Graham:2010hh}.  We leave further exploration of these signatures to future work, and here focus on assessing the relative likelihood of this possibility.

One way to approach this problem is to compare the rate of classical transitions to our vacuum to the rate of bubble nucleations of our vacuum.  The results of Section \ref{sec:coll} allow us to go a step further, since there we have computed the physical volume on a fixed scale-factor time hypersurface in the causal futures of these events.  In short, these are given respectively by the fourth and second terms on the right-hand side of the physical-volume rate equation, (\ref{rate2}).  Using $V_i\propto s_i\,e^{(3-q)t}$, we write the ratio of the volume in $i$ in the causal future of a classical transition to the volume in $i$ in the causal future of a bubble nucleation
\beq
R = \frac{\sum_{j,k,\ell}\lambda_{ik\ell j}\kappa_{kj}\kappa_{\ell j}(H_j/H_i)^2s_j}{\sum_j\kappa_{ij}s_j} \,.
\eeq    
Evidently, $R$ depends on the detailed microphysics of the landscape.  In particular, while the sums over vacua $j$ in the numerator and denominator both cover all dS vacua $j$, for almost all of these vacua the transition rates $\kappa_{ij}$ are zero, while the vacua $j$ for which the rates are nonzero are in general different in the numerator and denominator.  Since the components of the dominant eigenvector $s_i$ are exponentially staggered, $R$ could be enormous or zero depending on the detailed structure of a large set of transition rates in the landscape (zero corresponds to the case where classical transitions cannot create vacuum $i$). 

It is amusing to consider a highly idealized situation.  Suppose that classical transitions to our vacuum can only be caused by bubble collisions in one type of vacuum, and suppose that bubbles of our vacuum can only nucleate in one type of vacuum, and suppose that the Hubble rates and the relevant decay rates and the components of the dominant eigenvector for these vacua are the same: $H_0$, $\Gamma$, and $s_0$ respectively.  Finally, suppose that the classical transition converts the entire causal future of the collision into vacuum $i$.  In this situation, (\ref{gconstraint}) and (\ref{gdef}) imply that
$\sum_{k,\ell} \lambda_{ik\ell j} = 1$.  Putting all of this together, we find
\beq
R = \frac{4\pi}{3}\frac{H_0^2}{H_i^2}\frac{\Gamma}{H_0^4} \,.
\label{Rresult}
\eeq
This expression is familiar from the bubble-collision literature, and gives the typical number of bubble collisions in an observer's past lightcone (see for example \cite{Salem:2011qz}).  In this context, $H_i$ is the Hubble rate during slow-roll inflation in the observer's bubble, while $H_0$ and $\Gamma$ are respectively the Hubble rate and vacuum decay rate outside of the bubble.  The congruence between these two results is not surprising: the idealized situation that we have established above is equivalent to comparing the volume in the causal future of bubble collisions to the volume not in the causal future of bubble collisions, assuming that the volume in the causal future of two collisions is double-counted, etc.  This is equal to the number of bubble collisions in an observer's past lightcone.  The appearance of the inflationary Hubble rate $H_i$ in the bubble-collision result, compared to what one might suggest should be the Hubble rate associated with late-time cosmological constant domination in the relative probability (\ref{Rresult}), is due to the former focusing on observers like us who arise at finite time while the latter (interpreted as above) implicitly focuses on hypothetical observers at future infinity.

\section{Conclusions}
\label{sec:conclusions}

We have updated the rate equation describing the volume fractions of vacua in the multiverse to account for bubble collisions.  The rate equation refers to a global time foliation, and we have presented our analysis in terms of a scale-factor-time foliation, also providing a couple of brief remarks on how to translate the results into those of a lightcone-time foliation.  

As in the case where bubble collisions are ignored, the intuition for the asymptotic attractor solution to the rate equation revolves around the dS vacuum with the smallest decay rate.  Owing to its stability, this vacuum dominates the volume fraction of the multiverse---hence it is called the dominant vacuum---and the relative volume fractions of any other vacua are understood in terms of the relative likelihoods of the sequences of transitions that connect the dominant vacuum to them.  The most significant effect of bubble collisions stems from the possibility for classical transitions to vacua that are otherwise not involved in the collision.  These transitions should be included along with bubble nucleations in the sequences of transitions referred to above.  Thus, the existence of classical transitions modifies the detailed hierarchical structure among components of the dominant eigenvector.  

Although the volume fractions of many vacua might be exponentially larger on account of including classical transitions, the volume fractions of all dS vacua remain exponentially suppressed relative to the volume fraction of the dS vacuum with the smallest decay rate.  Thus, the notion of this vacuum as the dominant vacuum is maintained.  On the other hand, any modifications to the detailed hierarchical structure among components of the dominant eigenvector have relevance to the staggering and Boltzmann-brain issues.  A conclusive investigation of these issues is beyond the scope of this paper, yet we have discussed plausible circumstances under which accounting for classical transitions does not reveal a Boltzmann-brain problem.  In particular, Boltzmann-brain domination occurs only if the dominant vacuum has a sufficiently small vacuum energy and can set up classical transitions to a dS vacuum that supports Boltzmann brains but not to any dS vacua that form normal observers.

We have also explored the likelihood that our local Hubble volume was established by a classical transition, as opposed to a semiclassical bubble formation.  We found that the relative likelihood of these possibilities depends on the detailed relationships among transition rates in the landscape.

\acknowledgments

The author thanks Adam Brown, I-Sheng Yang, and Claire Zukowski for valuable discussions.  The author is also grateful for the support of the Stanford Institute for Theoretical Physics.

\providecommand{\href}[2]{#2}\begingroup\raggedright\endgroup

\end{document}